\documentclass[aps,prb,twocolumn,superscriptaddress, show pacs]{revtex4}

\usepackage{gensymb}
\usepackage{amssymb}
\usepackage{bm, amsmath}
\usepackage{bm}
\usepackage{physics}
\usepackage[euler]{textgreek}
\usepackage{siunitx}
\usepackage[hidelinks]{hyperref}
\usepackage[version=4]{mhchem}
\usepackage{natbib}

\usepackage{graphicx}
\usepackage[caption=false]{subfig}
\usepackage[outdir=./]{epstopdf}

\usepackage{booktabs}
\usepackage{dcolumn}
\usepackage{multirow}

\usepackage[normalem]{ulem} 
\usepackage{verbatim}
\usepackage[draft]{todonotes}
\usepackage{float}

\newcommand{\MCGa}{\ce{Mn_{0.8}CoGe}}
\newcommand{\MCGb}{\ce{Mn_{1.4}CoGe}}
\newcommand{\MCGc}{\ce{Mn_{2.5}CoGe}}
\newcommand{\MCGy}{\ce{Mn_{0.9}Co_{0.8}Ge}}
\newcommand{\MCGz}{\ce{Mn_{1.8}Co_{0.8}Ge}}

\newcommand{\numero}{N\textsuperscript{\underline{o} }}
\newcommand{\Pstmmc}{\ensuremath{P6_3/mmc}}

\newcommand{\TC}{\ensuremath{T_\text{C}}}
\newcommand{\HC}{\ensuremath{H_\text{C}}}
\newcommand{\MR}{\ensuremath{M_\text{R}}}
\newcommand{\MS}{\ensuremath{M_\text{S}}}
\newcommand{\Hext}{\ensuremath{\mu_{0}H_\text{ext}}}

\DeclareSIUnit{\Torr}{Torr}
\DeclareSIUnit{\bohr}{\mu_B}
\DeclareSIUnit{\fu}{f.u.}

\begin{document}
	
	\title{Disorder-induced ferrimagnetism in sputtered \ce{Mn_{x}CoGe} thin films}
	
	\author{D.~Kalliecharan}
	\author{J.~S.~R.~McCoombs}
	\author{M.~M.~E.~Cormier}
	\author{B.~D.~MacNeil}
	\author{R.~L.~C.~Molino}
	\author{T.~L.~Monchesky}\thanks{tmonches@dal.ca}
	\affiliation{Department of Physics and Atmospheric Science, Dalhousie University, Halifax, Nova Scotia, Canada B3H 3J5}
	\date{March 5, 2023}
	
	\begin{abstract}
		{
			Investigations into the magnetic properties of sputtered \ce{Mn_{x}CoGe} films in the range $0.8 \leq x \leq 2.5$ uncovered ferrimagnetic order, unlike the ferromagnetic order reported in bulk samples. These films formed hexagonal \ce{Ni_{2}In}-type structures when annealed at temperatures below \SI{600}{\degreeCelsius}. While the Curie temperatures of the films are comparable to those of hexagonal bulk \ce{MnCoGe}, there is a reduction in the magnetization of the \ce{Mn_{x}CoGe} films relative to bulk \ce{MnCoGe}, and a magnetization compensation point is observed in the $x<1$ samples. To understand the behavior, we calculated the magnetic moments of Mn-antisite defects in \ce{MnCoGe} with density-function theory (DFT) calculations. Models constructed from the calculation suggest that films become ferrimagnetic due to the presence of Mn on the Co and Ge sites. In the $x<1$ samples, these defects arose from the disorder in the films, whereas for $x>1$, the excess Mn was driven onto the antisites. Mean field modeling of the temperature dependence of the magnetization provides additional evidence for ferrimagnetism. Our mean field and DFT models provide a description of how the variation in film defects with composition will transition the magnetic behavior from a compensated (V-type) to an uncompensated (Q-type) ferrimagnet.
		}
	\end{abstract}
	
	
	\maketitle
	
	\section{Introduction}
	The manganese germanides comprise a rich phase diagram with a diverse range of magnetic structures. \ce{Mn_{3}Ge} forms one of two polytypes. The \ce{Mn_{3}Ge} $D0_{19}$ hexagonal structure is a frustrated non-collinear antiferromagnet with a large topological Hall effect,\cite{Nayak:2016sa} while the tetragonal $D0_{22}$ Heusler is a high-anisotropy ferrimagnet of interest for memory applications.\cite{Kurt:2012apl} There have been recent proposals for tuning the magnetic properties of this structure via chemical substitutions \ce{Mn_{3-y}X_{y}Ge}.\cite{You:2017jmmm} Substitution of Ni, for example, decreases the moment and increases the coercivity. \cite{Balluff:2018prb}
	
	A related family of compounds -- the inverse tetragonal Heuslers -- is obtained by replacing Mn on one of the 4d Wyckoff sites in the $D0_{22}$ structure, (0,~1/2,~1/4), with another element. This lowers the symmetry from $D_{3h}$ to the non-centrosymmetric $D_{2d}$ point group and turns on the Dzyaloshinskii-Moriya interaction that is responsible for the non-collinear magnetic structures in \ce{Mn_{2}RhSn} \cite{Meshcheriakova:2014prl} and \ce{Mn_{1.4}PtSn}. \cite{Nayak:2017nat}  The stability of the \ce{Mn_{2}XGe} Heusler compounds have been explored by density-functional theory (DFT) calculations,\cite{Faleev:2017pra} many of which are predicted to form the inverse tetragonal structure, including \ce{Mn_{2}CoGe}. The initial motivation for the work in this paper was to create \ce{Mn_{2}CoGe} Heusler alloy films by magnetron sputtering. We fabricated \ce{Mn_{x}CoGe} in the compositional range $0.8 \leq x \leq 2.5$, but were unsuccessful in producing Heusler alloys. The entire composition formed either a hexagonal structure or an orthorhomic structure related to the magnetocaloric material, \ce{MnCoGe}. This paper reports on alloys which formed the hexagonal structure.
	
	At low temperature, \ce{MnCoGe} forms an orthorhombic $C_{23}$ \ce{TiNiSi}-type structure (space group \numero 62, $Pnma$). It is a collinear ferromagnet with a Curie temperature, $T_C^\textrm{ortho} = $ \SI{355}{\kelvin} and a magnetic moment of {$m~=~$\SI{3.86}{\bohr}/formula unit} (f.u.). At a temperature $T_t$, the material undergoes a martensitic transformation to a hexagonal $B8_2$ \ce{Ni_{2}In}-type structure (space group \numero 194 \Pstmmc{}).\cite{Johnson:1975ic}  The resulting 3.9\% contraction in volume leads to a broadening of the Mn d-bands producing a smaller moment and lower $T_C$.\cite{Kaprzyk:1990jmmm}  In this hexagonal polytype, $m = $ \SI{2.78}{\bohr/\fu} \cite{Kanomata:1999jmsj} and $T_C^{\text{hex}} \approx $\SI{260}{\kelvin}. The martensitic transition is very sensitive to defects. Johnson \emph{et al.} found that $T_t$ varied between \SIrange{398}{453}{\kelvin},\cite{Johnson:1975ic}  while Kanomata \emph{et al.} reported $T_t$ as high as \SI{650}{\kelvin}.\cite{Kanomata:1999jmsj}  When $T_t$ lies between $T_C^{\textrm{hex}}$ and $T_C^\textrm{ortho}$ the material undergoes a first-order transition from an orthogonal ferromagnet to a hexagonal paramagnet that gives rise to a large magnetocaloric effect.
	
	\begin{figure}[!h]
		\centering
		\includegraphics[width=0.65\columnwidth]{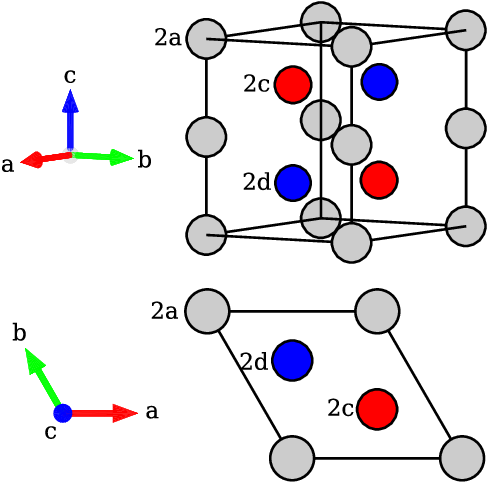}
		\caption{(Color online) The primitive unit cell of the \ce{MnCoGe} \ce{Ni_{2}In}-type phase. Isometric view (top) and c-axis projection (bottom), showing the 2a-Mn sites (grey), the 2c-Co sites (red) and 2d-Ge sites (blue).}
		\label{fig:xtal}
	\end{figure}

	What makes \ce{MnCoGe} particularly attractive is that its martensitic temperature can be chemically tuned independent of $T_C$. The transition temperature $T_t$ is very sensitive to Co vacancies, \cite{Kanomata:1995jmmm,Fang:2009jmmm} as well as Mn vacancies.\cite{Liu:2010epl}  With only a few percent vacancies on either site, $T_t$ can be reduced to room temperature with little effect on either $T_C^\textrm{hex}$ or $T_C^\textrm{ortho}$. This is potentially driven by a reduction in the number of valence electrons, as the same effect is also observed in \ce{Mn_{1+x}Co_{1-x}Ge} alloys.\cite{Ma:2011cpb, Liu:2010epl} Numerous studies have explored the influence of other defects and substitutions in \ce{MnCoGe}; a comprehensive summary of such studies is given in the appendix of Ref.~\onlinecite{Ren:2016phd}.
	
	In the \ce{Ni_{2}In}-phase, Mn resides on the 2a (0,~0,~0) Wyckoff sites and forms low density (001) planes. These are separated by dense \ce{CoGe} planes with Co on the  2c (1/3,~2/3,~1/4) sites and Ge on the 2d (2/3,~1/3,~1/4) sites (see Fig. \ref{fig:xtal}).
	
	We found that \ce{Mn_{x}CoGe} films prepared by DC magnetron sputtering were much more disordered than typical bulk material, which had two important consequences. Firstly, the hexagonal B8$_2$ phase was obtained at room temperature after annealing at $T = $ \SI{500}{\degreeCelsius} and remained in this phase upon cycling down to low temperature, consistent with other reports of sputtered \ce{MnCoGe} films.\cite{Portavoce:2018ass}  Secondly, the films display ferrimagnetic rather than ferromagnetic order reported in other investigations of this material.  We support the analysis of the magnetic properties with DFT calculations that show the spins from Mn-antisite defects align in the opposite direction to the spins on the Mn-sites.
	
	\section{Growth} \label{sec:expprep}
	Films were deposited on thermally oxidized Si wafers, as \ce{SiO_{2}} acts as a diffusion barrier for Mn, Co and Ge. \cite{takamura2008}  Si(001) wafers (manufactured by \emph{Prolog semicor Ltd.}) were cut into \SI{20}{mm} $\times$ \SI{20}{mm} squares and were sonicated in acetone and methanol baths for 15 minutes each. Before removing the wafers from the methanol bath, de-ionized nanopure water was slowly added and allowed to overflow in order to remove any contaminants from the liquid surface. The wafers were heated in a dry furnace at \SI{900}{\degreeCelsius} for 5 hours to create a \ce{SiO_{2}} layer, approximately \SI{300}{nm} in thickness.
	
	The sonication treatment was then repeated prior to loading samples into a \emph{Corona Vacuum Coater V3T} magnetron sputtering deposition system with a base pressure of \SI{3.0e-7}{\Torr}. The Ar pressure during sputtering was \SI{2.0e-3}{\Torr}. Films were deposited at room temperature, with no external heating. Sputtering rates were calibrated by measuring the weights of the samples before and after growth. The Mn sputtering rate was fixed at \SI{8.61}{\nano\mol\per\centi\metre\squared\per\second}, while the Co and Ge rates ranged from \SIrange{4.13}{12.9}{\nano\mol\per\centi\metre\squared\per\second}, depending on the stoichiometry. Film thickness was measured using a \emph{Vecco Dektak contact profilometer}. All films studied in this work were between \SI{475}{\nano\metre} and \SI{550}{\nano\metre} in thickness. The compositions were verified using a \emph{Thermo iCAP Q} laser ablation inductively coupled plasma mass spectrometer (LA-ICP-MS). The results are shown in Table~\ref{tbl:xrdparam}.
	
	The as-grown films where crystallized \emph{ex-situ} by annealing in an Ar environment in a \emph{Modular Process Technology RTP600s} Rapid Thermal Annealer (RTA). The RTA reached the desired temperatures within \SI{20}{\s} (\SIrange{15}{35}{\degreeCelsius\per\s}), and were cooled at a rate of approximately \SI{2}{\degreeCelsius\per\s}. An X-ray photoelectron spectroscopy (XPS) depth scan was performed on selected annealed samples, revealing that oxide contamination only exists at the surface, within the top 2\% of the film thickness.
	
	\section{Structural Characterization}
	The crystal structures of the films were investigated with conventional X-ray diffraction (XRD) $\theta-2\theta$ measurements on a \emph{Siemens D500 Diffractometer} equipped with a Cu source and monochrometer. To determine the strain in the films, the XRD measurements were compared to grazing angle X-ray diffraction (GAXRD) measurements, where the incident X-ray beam is fixed at $\theta_i = \ang{6}$. The alignment of the diffractometer was checked with a Si powder sample for both the XRD and the GAXRD geometries.
	
\begin{figure}[!h]
		\includegraphics{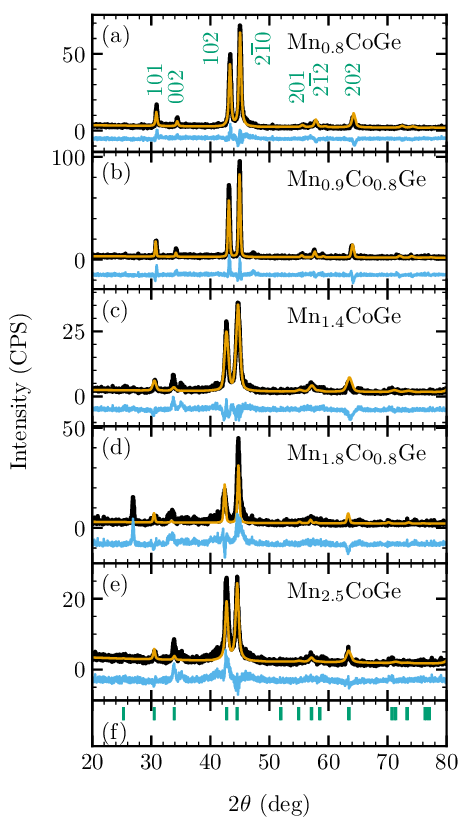}
		\caption{(Color online) XRD data (black) with Reitveld refinements (orange) and residuals (blue): (a) \MCGa{}, (b) \MCGy{}, (c) \MCGb{}, (d) \MCGz{} and (e) \MCGc{}, with \ce{Ni_{2}In}-type peak locations (green) in (f).}
		\label{fig:xrd}
	\end{figure}

	As-deposited XRD data shows that the films are either nanocrystalline or amorphous and discernible crystallographic phases only appeared after annealing. Annealing times and temperatures were selected to produce single phase samples. Five sets of samples --\MCGa{}, \MCGy{}, \MCGb{}, \MCGz{} and \MCGc{} -- were annealed within the temperature range of \SIrange{375}{600}{\degreeCelsius} for times between 2 minutes and 40 minutes, yielding \ce{Ni_{2}In}-type polycrystalline films. High temperature annealing resulted in mixed phase samples: annealing at \SI{700}{\degreeCelsius} produced a mixture of the hexagonal \ce{Ni_{2}In}-type and the orthorhombic \ce{TiNiSi}-type phases. The properties of the samples annealed at high temperature are not discussed further. Figure~\ref{fig:xrd} shows fits to GAXRD measurements of the \ce{Ni_{2}In}-type samples that demonstrate the phase is stable across the entire composition range, $0.8 \leq x \leq 2.5$.
	
\begin{figure}[!ht]
		\centering
		\includegraphics[width=0.85\columnwidth]{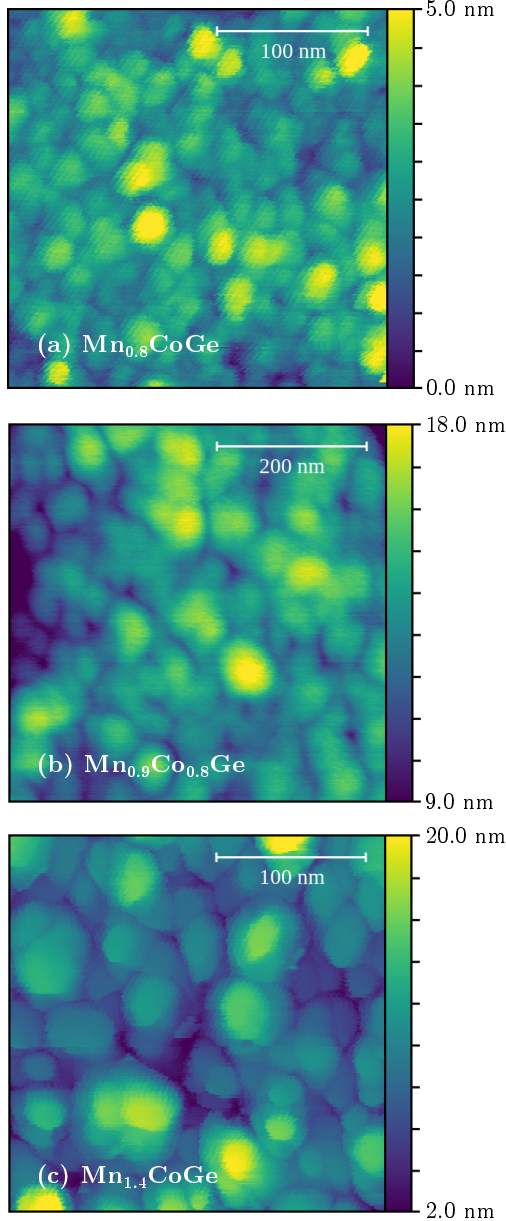}
		\caption{(Color online) AFM images of (a) \MCGa{}, (b) \MCGy{} and (c) \MCGb{}. The average grain sizes calculated via autocorrelation for (a) and (b) were \SI{56}{\nano\metre} and \SI{84}{\nano\metre}, respectively. The average grain size was calculated by inspection of the micrograph for (c) and was found to be \SI{23}{\nano\metre}.}
		\label{fig:afm}
	\end{figure}

	Estimates of the grain size were calculated from the diffraction peak widths (Fig.~\ref{fig:xrd}) by using the Scherrer equation:\cite{Scherrer1918}

	\begin{equation}
		\tau = \frac{K\lambda}{\beta\cos\theta}\,,
		\label{eq:scherrer}
	\end{equation}
	where $\tau$ is the grain size, $\lambda$ is the X-ray wavelength, and $\beta$ is the full-width at half-maximum of the diffraction peak at a Bragg angle $\theta$. The Scherrer constant $K$ is the crystallite-shape factor, chosen to be 0.9 for these samples. For each stoichiometry, the XRD grain size was averaged over several peaks and both Cu K$\alpha_1$ and Cu K$\alpha_2$ contributions to the peak were considered in determining $\beta$. The average grain sizes measured by XRD are summarized Table~\ref{tbl:xrdparam}. Grain sizes determined by atomic force microscopy (AFM) were largely in agreement with these estimates. Figure \ref{fig:afm} show representative micrograms of the \MCGa{}, \MCGy{} and \MCGb{} samples. The average grain diameter was taken as the first minimum in the autocorrelation of the height, $(h(\mathbf{r}) - h(\mathbf{r}_0))^2$. The extracted diameters for the \MCGa{}, \MCGy{} samples were \SI{56}{\nano\metre} and \SI{84}{\nano\metre}, respectively, in agreement with the XRD estimates shown in Table \ref{tbl:xrdparam}. For \MCGb{}, the autocorrelation function yielded a value of \SI{72}{\nano\metre}, nearly 3 times that from XRD. Figure \ref{fig:afm}(c) shows the presences of smaller features  on top of the larger 70 nm diameter grains that are 23 nm in diameter on average, which is within error of the grain size extracted from XRD. This suggests that the larger 70 nm features in \ref{fig:afm}(c) are in-fact composed of smaller crystallites.  
	
	The lattice parameters extracted from the GAXRD fits (Table~\ref{tbl:xrdparam}), are comparable to the values of bulk \ce{MnCoGe}, $a=4.087(1), c=5.316(3)$~\AA.\cite{Jeitschko:1975acb}  The Rietveld refinements were performed using Rietica version 4.0 (\mbox{\url{http://rietica.org}}). We note that the (101) peak intensity is much lower that expected from bulk \ce{MnCoGe} samples. The discrepancy could be accounted for with 20\% vacancies on the 2c-site occupied by Co. The presence of vacancies is supported by	 ICP-MS measurements that show Mn concentrations are lower than the nominal value. The intensity of the ($2\bar{1}0$)-peaks is higher than expected. As the annealing process can lead to preferred grain orientation, it is not possible to separate this effect from the possibility of vacancies.
	
	\begin{table}[!ht]
		\caption{The composition of the \ce{Mn_{x}CoGe} films determined by LA-ICP-MS, together with lattice parameters determined from GAXRD. Grain sizes determined via XRD (\textit{cf}. Eq.~\ref{eq:scherrer}) are also given, which agree with those found from AFM.}
		\label{tbl:xrdparam}
		\centering
		\bigskip
		\begin{tabular}{c c c c c c c}
			\hline\hline
			\textbf{\textit{x}} & $\boldsymbol{\chi}_\textbf{Mn}/\boldsymbol{\chi}_\textbf{Ge}$ & $\boldsymbol{\chi}_\textbf{Co}/\boldsymbol{\chi}_\textbf{Ge}$ & \textbf{\textit{a} (\AA)} & \textbf{\textit{c} (\AA)}   & $\boldsymbol{\tau}_\textbf{XRD} (\si{nm})$ \\ [0.25ex]
			\hline
			0.8 & 0.79 & 0.93 & 4.02 & 5.21  & 56.04 \\
			0.9 & 0.89 & 0.79 & 4.03 & 5.23  & 84.03 \\
			1.4 & 1.42 & 0.97 & 4.05 & 5.30  & 23.99 \\
			1.8 & 1.83 & 0.78 & 4.05 & 5.36  & 42.00 \\
			2.5 & 2.47 & 1.07 & 4.07 & 5.29  & 33.56 \\
			\hline\hline
		\end{tabular}
	\end{table}
	
\begin{figure}[!ht]
		\includegraphics[width=0.85\columnwidth]{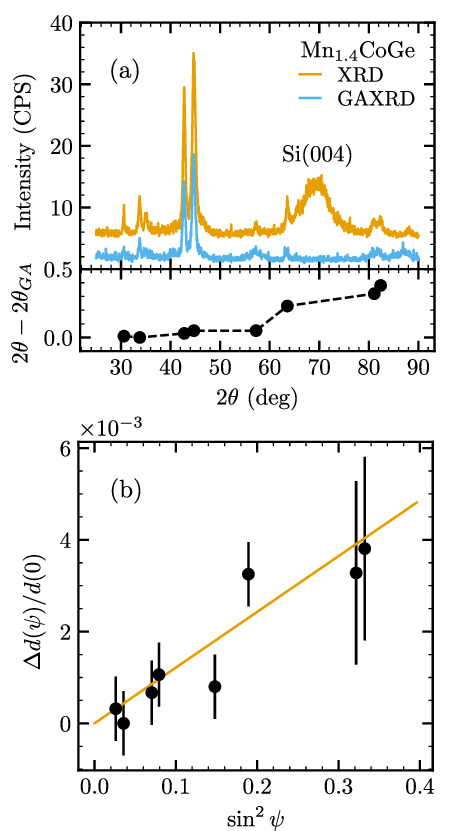}
		\caption{(Color online) (a) XRD and GAXRD measurements of \MCGb{}. The lower panel shows the XRD peak positions relative to GAXRD peaks. Note that the intensity of the Si(004) peak at $2\theta=\ang{69.9}$ was reduced by offsetting the sample angle by $\ang{2}$. (b)	The fractional change in the measured interatomic plane spacing. Data has been linearized and the solid line shows the fit to the data using Eq.~\eqref{eq:dpd}.}
		\label{fig:strain}
	\end{figure}

	The GAXRD peak positions were found to be systematically lower than the XRD measurements. This shift was not present in the control Si powder sample. A comparison between GAXRD and XRD is shown in Fig.~\ref{fig:strain}~(a). While XRD probes the lattice parameters of planes that are parallel to the substrate surface, GAXRD measures interatomic planes whose normal is further and further from the film normal as the detector angle $\theta$ increases. We define $\psi = \theta - \theta_i$ as the angle between the film's normal and the scattering vector. As we show, the shift in the GAXRD peaks relative to those in the conventional XRD measurements is due to strain in the films.
	
	To determine the influence of film strain on the GAXRD measurements, we assume a uniform biaxial strain of the polycrystalline material, where  $\epsilon_{\perp}$ and $\epsilon_{\|}$ are the out-of-plane and in-plane strains, respectively. The strain for planes that are at an angle $\psi$ with respect to the film surface is given by Eq. 13 in Ref.~\onlinecite{Welzel2005} for the case of zero shear strain,
	$\epsilon(\psi)~=~\epsilon_{\|}\sin(\psi)^2~+~\epsilon_{\perp}\cos(\psi)^2$,
	from which one obtains the ratio of planes spacing measured for the scattering vector along $\psi$ relative to those along $\psi = 0$  :
	\begin{equation}\label{eq:dpd}
		\frac{d(\psi)}{d(0)} = \left(\frac{1 + \epsilon_\parallel\sin^2\psi + \epsilon_\perp\cos^2\psi}{1 + \epsilon_\perp}\right).
	\end{equation}
	For small strain, Eq.~\ref{eq:dpd} can be written as $\Delta d(\psi)/d(0)~=~[d(\psi)~-~d(0)]/d(0)~\approx~(\epsilon_\parallel~-~\epsilon_\perp)\sin~\psi^2$,  which allows us to extract $\epsilon_\parallel - \epsilon_\perp = 0.012
	\pm0.01$ for the \ce{Mn_{1.4}CoGe} sample in Fig. \ref{fig:strain}(b).
	
	The strain, which was observed for all \ce{Mn_{x}CoGe} films, is likely induced by the annealing process. The thermal expansion coefficients for metals are typically about one order of magnitude larger than the Si substrate. The film crystallizes at high temperature; since the film contracts more than the substrate upon cooling, the film develops an in-plane tensile strain (and through the Poisson ratio, it develops an out-of-plane compressive strain).
	
	\section{Magnetic measurements}
	Magnetic measurements were performed using a \emph{Quantum Design Physical Properties Measurement System} (PPMS), equipped with a \emph{P500 AD/DC Magnetometry System} (ACMS). Samples were cut into {\SI{5.8}{mm} $\times$ \SI{5.8}{mm}} squares and wedged into a plastic straw that was placed in the PPMS. The field was applied in the plane of the film.
	
\begin{figure}[!ht]
		\centering
		\includegraphics{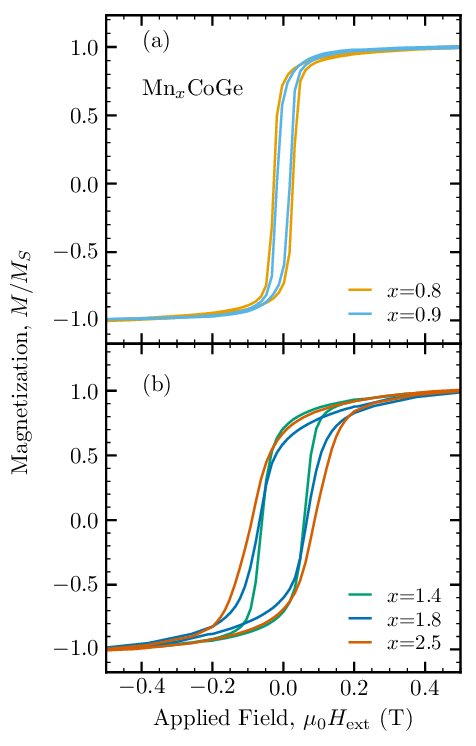}
		\caption{(Color online) Normalized hysteresis curves of \ce{Mn_{x}CoGe} films for compositions (a) $x<1$ and  (b) $x>1$ measured at $T=\SI{5}{\kelvin}$. The saturation magnetizations are given in Table~\ref{tbl:magparam}.}
		\label{fig:hyst}
	\end{figure}

	Magnetization loops were recorded as the field was cycled between $\mu_0 H = \pm\SI{9}{\tesla}$. The $M-H$ loops for all five samples measured at $T = $ \SI{5}{\kelvin} are qualitatively similar, as shown in Fig.~\ref{fig:hyst}.  However, hysteresis loops with $x>1$ show larger \HC{} with more rounding, suggestive of a larger mean effective anisotropy with a broader distribution.
	
\begin{figure}[!h]
		\centering
		\includegraphics{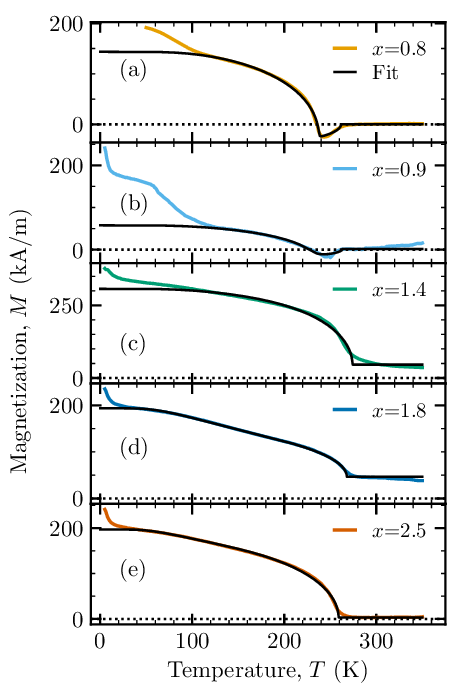}
		\caption{(Color online) Remanent magnetization vs temperature after field-cooling to $T = 5$~K.  The nominal structures \MCGa{} and \MCGy{} show V-type ferrimagnetic behavior due to Mn occupancies on 2a and 2c sites. \MCGb{}, \MCGz{} and \MCGc{} are Q-type ferrimagnets. \MCGb{} and \MCGz{} show a secondary magnetic phase. Fits are provided based on a two-sublattice ferrimagnetic model, described in Eq.~\ref{eq:ferrimag}.}
		\label{fig:M-T}
	\end{figure}

	The remanent magnetization, \MR{}, was measured on warming from $T = $ \SI{5}{\kelvin} after saturating the film in a \SI{9}{\tesla} field. The temperature dependence of \MR{} is shown in Fig.~\ref{fig:M-T}. Some of the samples with composition $x=2.5$ had a small remanent magnetization above $T = $ \SI{270}{\kelvin}. Although no impurity phase could be detected in the X-ray measurements, additional annealing in the RTA was able to remove this additional ferromagnetic contribution. The $x=1.4$ and $1.8$ samples also show a small \MR{} above $T = $ \SI{270}{\kelvin} but further annealing could not remove the impurity phase. Unexpectedly, the compositions with $x<1$ exhibited a distinctly ferrimagnetic behavior: above a compensation point of approximately $T = $ \SI{230}{\kelvin}, the \MR{} reverses sign. Though the $M_\text{R}-T$ curves for $x>1$ may resemble those of a ferromagnet, we will argue in subsequent sections that each sample exhibits an $M_\text{R}(T)$ curve consistent with Q-type or V-type ferrimagnetism.
	
	The Curie temperature is estimated from the knee in the $M_\text{R}-T$ plot as $M_\text{R}$ approaches zero. As shown in the Table~\ref{tbl:magparam}, \TC{} is comparable to the bulk $T_C^\text{hex} \approx$ \SI{260}{\kelvin} of the hexagonal phase, and is relatively insensitive to the composition $x$, as observed in bulk. \cite{Liu:2010epl}  However, the table also shows that the total magnetic moment per primitive unit cell is significantly lower that the bulk value for \ce{MnCoGe}, \SI{5.56}{\bohr} per primitive unit cell.
	
	\begin{table}[!h]
		\caption{The saturation magnetization, \MS, the magnetic moment per primitive unit cell, $m$, the coercive field \Hext{} and Curie Temperature $T_C$ for \ce{Mn_{x}CoGe} films.}
		\label{tbl:magparam}
		\centering
		\bigskip
		\begin{tabular}{c c c c c}
			\hline\hline
			\textbf{\textit{x}} & \textbf{\textit{M}\textsubscript{\text{s}} (kA/m)} & \textbf{\textit{m}} ($\boldsymbol{\mu_B}$) & \textbf{\textit{H}\textsubscript{C} (mT)} &\textbf{ \textit{T}\textsubscript{C} (K)} \\
			\hline
			0.8 & 380 & 2.99 & 26 & 267\\
			0.9 & 384 & 3.02 & 20 & 260\\
			1.4 & 497 & 4.19 & 78 & 277\\
			1.8 & 394 & 3.19 & 69 & 272\\
			2.5 & 353 & 2.87 & 89 & 267\\
			\hline\hline
		\end{tabular}
	\end{table}
	
	\section{Computed Magnetic Moments from Density-Functional Theory}
	
	To explore the origin of the drop in magnetic moment and the appearance of ferrimagnetic behavior, we considered the influence of atomic disorder in the \ce{Ni_{2}In} structure on the individual magnetic moments. In the ordered phase, nuclear magnetic resonance (NMR) shows that Mn on the 2a-site has a magnetic moment of  $m_\text{Mn} = $ \SI{2.4}{\bohr}, while the moment of Co on the 2c-site couples ferromagnetically to the 2a-site with a moment $m_\text{Co} = $ \SI{0.4}{\bohr}.\cite{Kanomata:1999jmsj}  These values are in good agreement with the measured magnetization and consistent with neutron scattering experiments. \cite{Kaprzyk1990} However, we note that DFT overestimates the magnetic moment of Mn in \ce{MnCoGe}, \cite{Kanomata:1999jmsj, Kaprzyk:1990jmmm, Hahn2017} and so has to be rescaled to compare to experiment.
	
	Previously published DFT calculations of \ce{Ni_{2}In}-type \ce{Mn_{2}Ge}  predict ferrimagnetic behavior due to the anti-parallel coupling between Mn on 2a- and 2c-sites. \cite{Ellner:1980jac} This is consistent with tight-binding (TB) calculations for \ce{MnCoGe} that show a reduction in the average Mn moment when it is distributed on both of these sites. The TB calculations show that Co on the other hand is little affected by either moving it to the 2a-site, or by the presence of Mn-antisite defects, as supported by DFT calculations.\cite{Hahn2017} However, there are very few studies of the \ce{Ni_{2}In}-type structure and the magnetic behavior of Mn on the 2c- and 2d-sites (Mn$_{2c}$ and Mn$_{2d}$) remains unclear.
	
	DFT \cite{Hohenberg1964, Kohn1965} computations were performed within the spin-polarized general gradient approximation (GGA) \cite{Perdew1996} using the Vienna \textit{Ab-initio} Simulation Package (VASP).\cite{Kresse1993,Kresse1996,Kresse1999,Kresse2007}
	Local magnetizations are obtained by projecting the ground state crystal orbitals onto atomic-like orbitals centered at each crystallographic site (\textit{i.e.} atom-centered).
	Since the magnetization can be strongly dependent on the inter-atomic distances, full cell relaxations were performed for all structures, converging forces to better than \SI{10}{\milli\eV/\angstrom} and stresses to within \SI{1}{\mega\pascal} by enforcing a sufficiently dense $k$-point sampling of the first Brillouin zone. We used projector augmented wave (PAW) datasets with 7, 9, and 4 valence electrons for Mn, Co, and Ge, respectively. The ground state energies were converged to better than \SI{1}{\milli\eV/\fu} using a plane-wave energy cut-off of \SI{550}{\eV}. We attempted to converge both ferromagnetic and ferrimagnetic solutions for all structures. In some cases both solutions converged, but we present here only the lowest energy solutions.
	
	Our computed magnetic moments for \ce{Ni_{2}In}-type \ce{MnCoGe} and \ce{Mn_{2}Ge} agree well with previously calculated values. In \ce{MnCoGe}, our computations show a slightly smaller Mn moment, \SI{2.75}{\bohr} compared to the value calculated in Ref.~\onlinecite{Hahn2017} (\SI{3.09}{\bohr}), but one that is closer to the experimental value. We obtain a moment of \SI{0.5}{\bohr} on Co, and \SI{-0.1}{\bohr} on Ge that are in good agreement with Ref.~\onlinecite{Hahn2017}, as well as experimental values.  In \ce{Mn_{2}Ge}, our computed magnetizations for Mn$_{2a}$, \SI{2.9}{\bohr} and Mn$_{2c}$, \SI{-2.0}{\bohr}, agree exactly with previously published DFT results. \cite{Arras:2011prb}. Unlike what has been published previously, we find that the ferrimagnetic state is not the ground states of the systems:  a spin-configuration with ferromagnetically aligned spins on the 2a-sites in the (001) plane but with antiferromagnetic alignment between neighboring (001) planes and zero moment on the 2c-sites results in a lower energy state. However, given that antiferromagnetism is not observed in any of the samples, the moments in the ferrimagnetic state of \ce{Mn_{2}Ge} provides a better reference for the spins in our samples and are used in the discussion below. 
	
	\begin{figure}[!h]
		\centering
		\includegraphics[width=0.85\columnwidth]{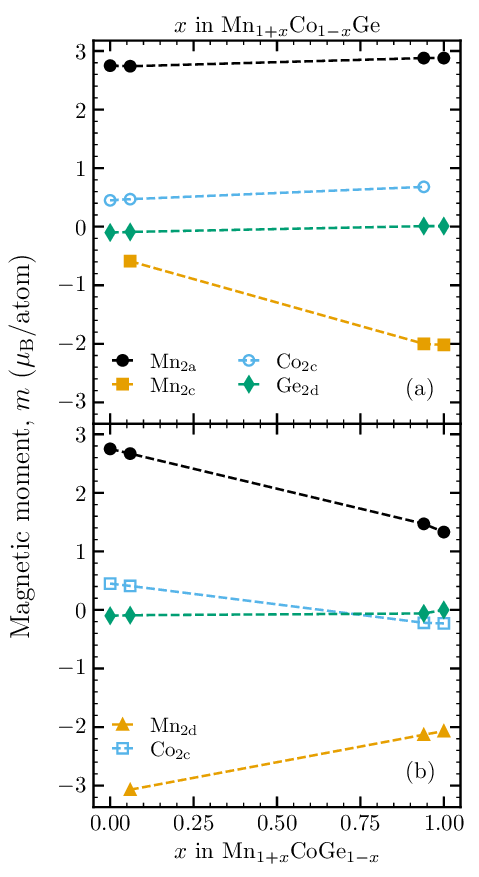}
		\caption{(Color online) DFT computed moments for \ce{Mn_{1+y}Co_{1-y}Ge} (top) and \ce{Mn_{1+y}CoGe_{1-y}} (bottom). The dotted lines represent linear interpolations used to model the experimental data.
		}
		\label{fig:dftmag}
	\end{figure}

	To determine the effect of Mn$_{2c}$, $2\times2\times2$ supercells were built by repeating the \ce{MnCoGe} hexagonal unit cell (6 atoms) twice along each lattice vector resulting in 16 Mn, 16 Co, and 16 Ge atoms. We considered the \ce{Mn_{1+y}Co_{1-y}Ge} solid solution where the excess Mn, $y$, replaces Co on the 2c site. For the dilute limit we placed 1 Mn on the 2c-site per supercell ($y = 0.06$); in the concentrated limit 15 of the 16 2c-sites were occupied by Mn ($y = 0.94$),  We note that the case of $y=0$ and $y=1$ correspond to \ce{MnCoGe} and \ce{Mn_{2}Ge}. The results are shown in Fig.~\ref{fig:dftmag}(a). The influence of Mn substitution onto the 2d site with an analogous  \ce{Mn_{1+y}CoGe_{1-y}} solid solution is shown in Fig.~\ref{fig:dftmag}(b).
	
	The Mn$_\text{2c}$ has little impact on the magnetic moments of either the Mn$_{2a}$ moments or the Co or Ge moments. However Mn$_{2c}$ does have a significant compositional dependence and is antiferromagnetically coupled to the Mn$_{2a}$ moments. In the dilute limit, the Mn$_{2c}$ moment of \SI{-0.6}{\bohr} is opposite in sign but comparable in magnitude to the Co moment. The magnetic moment of Mn$_{2c}$ reached \SI{-1.86}{\bohr} in the concentrated Mn$_{2c}$ regime, which approaches the calculated value for \ce{Mn_{2}Ge}, as expected.
	
	Despite the identical symmetry of the 2c- and 2d-sites, Mn behaves very differently when it is substituted on the Ge-sites due to its magnetic Co neighbors in the (001) plane. In the dilute limit, its moment is slightly larger than the 2a-moment giving a total moment of $0.08~\mu_{\rm B}$/f.u. As the concentration $y$ increases the magnitude of the Mn$_{2a}$- and Mn$_{2c}$-moments both decrease and so the heavy compensation continues for larger concentrations. 
	
	We also performed additional DFT calculations to examine the influence of Mn on both the 2c- and 2d-sites. We replaced one Co and one Ge atom in the $2\times2\times2$ supercells with Mn to give \ce{Mn_{1.12}Co_{0.94}Ge_{0.94}}. Two different configurations of this stoichiometry were generated -- one where the Mn on the 2c-site was nearest to the Mn on the 2d-site, another where it was farthest. All three configurations resulted in the same magnetic moment of \SI{-2.9}{\bohr} for Mn on the 2d-site and an unchanged magnetic moment for Mn on both the 2a- and 2c-sites.
	
	\section{Discussion}
	
	To understand whether the antiferromagnetically aligned \ce{Mn_{2c}} moments can explain the observed reduction in the magnetization, we construct a model of the defect distribution in the unit cell and use DFT calculations to estimate the magnetic moments.  Although DFT and the measured compositions differ somewhat in the amounts of Co and Ge, the magnetization is dominated by the size of the Mn moments. Therefore we use the moments calculated for \ce{Mn_{1+y}Co_{1-y}Ge} and \ce{Mn_{1+y}CoGe_{1-y}} that correspond to the same concentration of Mn in \ce{Mn_{x}CoGe}, given by $y = 2(x-1) / (x + 2)$. 
	
	Based on the X-ray analysis, we consider the possibility of vacancies on the 2c-sites, which changes  the relative number of 2c-sites relative to the 2a- and 2d-sites. In a sample that contains $f$ formula units of Mn$_x$CoGe, there are $n = f(x+2)$ atoms. However, in the presence of $n\nu$ vacancies on the 2c-sites, the  $n$ atoms require a total of  $n(1 + \nu)$, crystallographic sites. When filling these sites, we need to distinguish compositions according to the number 2a-sites relative to the number of Mn atoms. For the case where the difference between the number of Mn atom and the number of 2a-sites, ${\Delta_{\textrm{Mn}} = nx/(x+2) - n(1 + \nu)/3}$, is greater than zero, our model assumes that the excess is distributed on the remaining sites according to the relative number of 2c- and 2d-sites. Therefore we place ${\Delta_{\textrm{Mn}}(1 - 2\nu)/ (2 - \nu)}$ Mn atoms on 2c, and ${\Delta_{\textrm{Mn}}(1 + \nu)/ (2 - \nu)}$ on 2d, as shown Table~\ref{tbl:ex_Mn>0} together with the Co and Ge distributions.
	
	In the case where $\Delta_{\textrm{Mn}} < 0$, all of the Mn is accommodated on the 2a sites, and the remaining 2a-sites are filled by $|\Delta_{\textrm{Mn}}|/2$ Co atoms and $|\Delta_{\textrm{Mn}}|/2$ Ge atoms. Table \ref{tbl:ex_Mn<0}  shows the distribution of atoms for this case.
	
	We require additional site disorder to account for the reduction in the magnetization observed in our films. We considered both disorder between the 2a- and 2c-sites, as well as between the 2a- and 2d-sites. While both models can explain the size of the moments in our \ce{Mn_xCoGe} samples, the 2a-2d site disorder is required to explain the mean-field results described below. We therefore introduce a parameter $\delta$ that characterizes the fraction of the Mn$_{2a}$ that is exchanged with Ge$_{2d}$. 
	
	\renewcommand{\arraystretch}{2}
	\begin{table}[]
	  \centering
		\caption{The distribution of atoms for \ce{Mn_xCoGe} relative to the (x+2) atoms in the formula unit , for the case $\Delta_{\textrm{Mn}} \geq 0$. The fraction of Mn that is in excess of the available 2a sites is $\Delta_{\rm Mn} = x/(x+2) - (1 + \nu)/3$. $\nu$ is the number of vacancies per (x+2) atoms.} 
		\label{tbl:ex_Mn>0}
		\begin{tabular}{l|ccc|}
			\multicolumn{1}{r}{} & \multicolumn{1}{c}{\textbf{Mn}} & \multicolumn{1}{c}{\textbf{Co}} & \multicolumn{1}{c}{\textbf{Ge}} \\
			\cline{2-4}
			\multicolumn{1}{r|}{\textbf{2a}} & \multicolumn{1}{c|}{$\frac{1 + \nu}{3} - \frac{\delta(1+\nu)}{3}$}                            & \multicolumn{1}{c|}{0}                                                                                   & $\frac{\delta(1+\nu)}{3}$                   \\
			\multicolumn{1}{r|}{\textbf{2c}} & \multicolumn{1}{c|}{$\Delta_{\textrm{Mn}} \frac{1 - 2 \nu}{2 - \nu}$}        & \multicolumn{1}{c|}{$\frac{1 - 2 \nu}{3}- \Delta_{\textrm{Mn}} \frac{1 - 2 \nu}{2 - \nu}$}            & 0                          \\
			\multicolumn{1}{r|}{\textbf{2d}} & \multicolumn{1}{c|}{$\Delta_{\textrm{Mn}} \frac{1 + \nu}{2 - \nu} + \frac{\delta(1+\nu)}{3}$} & \multicolumn{1}{c|}{$\frac{1}{x + 2} - \frac{1 - 2\nu}{3} + \Delta_{\textrm{Mn}}\frac{1-2\nu}{2 - \nu}$} & $\frac{1}{x + 2} - \frac{\delta(1+\nu)}{3}$ \\
			\cline{2-4}
		\end{tabular}
	\end{table}
	
	\begin{table}[]
		\caption{The distribution of atoms for \ce{Mn_xCoGe} relative to the (x+2) atoms in the formula unit , for the case $\Delta_{\textrm{Mn}} < 0$.}
		\label{tbl:ex_Mn<0}
		\begin{tabular}{|l|ccc|}
			\multicolumn{1}{r}{} & \multicolumn{1}{c}{\textbf{Mn}} & \multicolumn{1}{c}{\textbf{Co}} & \multicolumn{1}{c}{\textbf{Ge}} \\
			\cline{2-4}
			\multicolumn{1}{r|}{\textbf{2a}} & \multicolumn{1}{c|}{$\frac{x}{x+2} - \frac{\delta(1+\nu)}{3}$} & \multicolumn{1}{c|}{$\frac{|\Delta_{\textrm{Mn}}|}{2}$}                                       & $\frac{|\Delta_{\textrm{Mn}}|}{2} + \frac{\delta(1+\nu)}{3}$                    \\
			\multicolumn{1}{r|}{\textbf{2c}} & \multicolumn{1}{c|}{0}                        & \multicolumn{1}{c|}{$\frac{2}{x+2} -|\Delta_{\textrm{Mn}}| - \frac{1 + \nu}{3} $}     & 0                                                      \\
			\multicolumn{1}{r|}{\textbf{2d}} & \multicolumn{1}{c|}{$\frac{\delta(1+\nu)}{3}$}                 & \multicolumn{1}{c|}{$\frac{1 + \nu}{3} - \frac{1}{x + 2} + \frac{|\Delta_{\textrm{Mn}}|}{2}$} & $\frac{1}{x + 2}  - \frac{|\Delta_{\textrm{Mn}}|}{2} - \frac{\delta(1+\nu)}{3}$ \\ 
			\cline{2-4}
		\end{tabular}
	\end{table}
	\renewcommand{\arraystretch}{1}
	
	\begin{figure}[!h]
		\centering
		\includegraphics{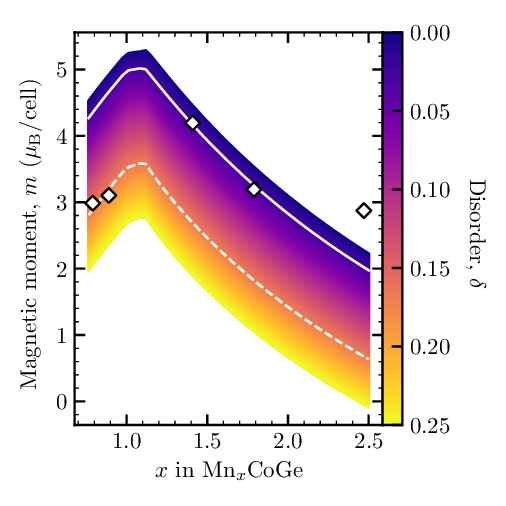}
		\caption{(Color online) The diamonds show the measured magnetic moment $m$ per primitive unit cell of \ce{Ni_{2}In}-type \ce{Mn_{x}CoGe} films. The color-plot shows the expected variation in the magnetic moment due to the disorder parameter $\delta$ with 20\% vacancies on the 2c-sites ($\nu = 0.07$). The solid and dashed lines show the calculated moment for $\delta = 0.03$, and 0.17 respectively.}
		\label{fig:disorder}
	\end{figure}
	
	To calculate the moments of the $\Delta_{\textrm{Mn}} \geq 0$ samples, we use the interpolated DFT moments shown by the dotted lines in Fig.~\ref{fig:dftmag}. We use the Mn$_{2c}$ and Mn$_{2d}$ occupancies obtained from Tables~\ref{tbl:ex_Mn>0} and \ref{tbl:ex_Mn<0} to determine the relative weights of the two sets of moments shown in Fig.~\ref{fig:dftmag}. Since DFT overestimates the Mn$_{2a}$ moment by a factor 2.75/2.4, we rescale all the predicted Mn moments by the corresponding amount. For the $\Delta_{\textrm{Mn}} < 0$ samples, DFT results in Ref.~\onlinecite{Hahn2017}  show that the \ce{Co_{2a}} and \ce{Ge_{2a}} antisite defects do not significantly affect the \ce{Mn_{2a}} moments and therefore use our $x=1$ calculated values.
	
	The calculated magnetic moment for 20\% vacancies on the 2c-sites ($3 \nu / (1 + \nu) = 0.2$) as a function of $x$ and $\delta$ is shown by the lines and color plot in Fig.~\ref{fig:disorder}. The peak in the plot occurs for $x = 1.11$, $\delta = 0$, corresponding to $\Delta_{\textrm{Mn}} = 0$, the maximum in the possible fraction of Mn on the 2a-sites. Below $\Delta_{\textrm{Mn}} = 0$, the modeled moment drops with decreasing $x$ due to a reduction in the available Mn. Above $\Delta_{\textrm{Mn}} \geq 0$, the moment drops with increasing $x$ as more Mn is forced onto the 2c-sites and 2d-sites. The color scale reflects the decrease in magnetic moment with increasing 2a-2d site disorder; a comparison with the data points allows an estimation of the disorder, $\delta$. The model suggest that the disorder could be as large as $\delta = 0.17$ for $\Delta_{\textrm{Mn}} < 0$ (dashed white line), and then drop below $\delta = 0.05$ for $\Delta_{\textrm{Mn}} < 0$ (solid white line). We note that the $x = 2.5$ sample  has a moment that is larger than can be explained by our model. One possible source for the discrepancy may be due to the inaccuracies of interpolating the DFT results. Nevertheless, the model captures the general trend in the variation of the saturation magnetization with Mn concentration. The model for the data and DFT results indicate that ferrimagnetism exists for all samples, not just the $\Delta_{\textrm{Mn}} < 0$ samples where compensated ferrimagnetic is observed. 
	
	Ferrimagnetism for the $\Delta_{\textrm{Mn}} > 0$ samples is not immediately obvious from the magnetometry measurements. However, a closer inspection of the shape of the $M-T$ plots in Fig.~\ref{fig:M-T} reveals features that are observed in other ferrimagnets, \cite{Smart1966} such as the linear $M_\text{R}(T)$  region in Fig.~\ref{fig:M-T}(d) between 80~K and 220~K. To explore the shape of the magnetization curves in more detail, we fitted the $M_\text{R}(T)$ curves with N\'eel's molecular field model.\cite{Neel1948} Since the DFT calculations show that Mn and Co behave similarly on the 2c-sites and the 2d-sites, we approximated the system with a two-sublattice model where $A$ refers to the moments on the 2a-sites and $B$ contains both the 2c- and 2d-sites. The molecular fields experienced by sublattice $A$ and $B$ are given by the usual mean-field parameters $\lambda_{ij}$,
	\begin{align}
		H_A(T) &= \lambda_{AA}M_{A}(T) + \lambda_{AB}M_{B}(T) \\ \nonumber
		H_B(T) &= \lambda_{AB}M_{A}(T) + \lambda_{BB}M_{B}(T).
	\end{align}
	The temperature-dependent magnetization of each sublattice is then calculated by solving the two coupled nonlinear equations,\cite{Neel1948}
	\begin{equation}
		\begin{aligned}
			M_{A}(T) &= M_A(0) \, 
			{\rm B}_{J_A}\!\qty(\frac{\mu_0m_{A} H_A(T)}{k_\text{B}T}) \, , \\
			M_{B}(T) &= M_B(0) \, 
			{\rm B}_{J_B}\!\qty(\frac{\mu_0m_{B} H_B(T)}{k_\text{B}T})\, ,
			\label{eq:ferrimag}
		\end{aligned}
	\end{equation}
	where ${\rm B}_{J_i}(y)$ is the Brillouin function. The molecular field coefficients are related to the exchange constant of the Heisenberg model of the form 
	\begin{equation}
	\mathcal{H} = -\sum\limits_{\left<i,j\right>}\mathcal{J}_{ij} \vb{S}_i \cdot \vb{S}_j
	\end{equation}
	through the relationship
	\begin{equation}
		\mathcal{J}_{ij} = \frac{\mu_0 (g \mu_\text{B})^2 \lambda_{ij}}{2z_{ij}V_\text{uc}} \,
	\end{equation}
	where $V_\text{uc} = \sqrt{3}a^2c/2$ is the unit cell volume and $z_{ij}$ is the number of $j$-sublattice nearest neighbours to atoms on sublattice $i$.

\begin{figure}[!h]
		\centering
		\includegraphics{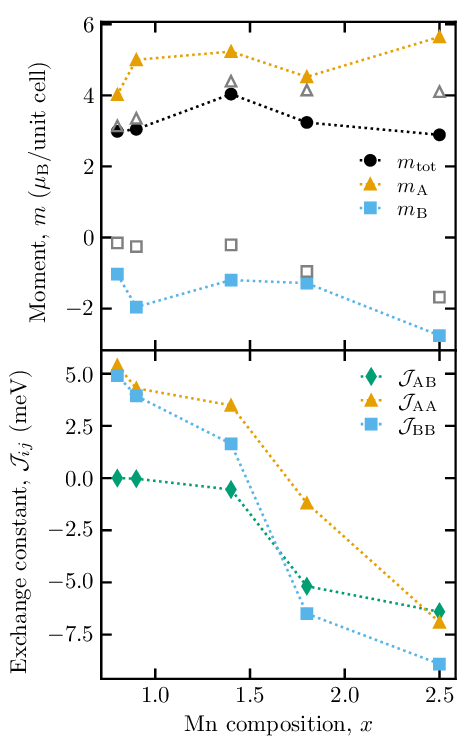}
		\caption{(Color online) Mean-field fitting parameters obtain from the fits in Fig.~\ref{fig:M-T} for sublattices A (2a-sites) and B (2c + 2d-sites). a) The magnetic moments per unit cell are shown by the filled colored points. For comparison, the open grey points show the moments from the the DFT-based model. b) The exchange constants for the inter-sublattice interaction $\mathcal{J}_{AB}$ and  the intra-sublattice interactions $\mathcal{J}_{AA}, \mathcal{J}_{BB}$.}
		\label{fig:Jex}
	\end{figure}

	We use the moments obtained from our DFT-based defect model, described by Table~\ref{tbl:ex_Mn>0} and \ref{tbl:ex_Mn<0}, as initial guesses for the mean-field sublattice moments ${m_{i} = g\mu_\text{B}\sqrt{J_i(J_i+1)}}$. The three molecular field coefficients $\lambda_{ij}$, together with the two sublattice moments $m_i$ are treated as fitting parameters. The resulting least-squares fits to the $M_\text{R}(T)$ data are shown by the black lines in Fig.~\ref{fig:M-T}, with the corresponding fitting parameters plotted in Fig.~\ref{fig:Jex}. It should be noted that attempts to fit the $x>1$ samples with Weiss' ferromagnetic mean field model were unsuccessful.  The fitted moments have been scaled to the saturation magnetizations listed in Table \ref{tbl:magparam}. 
	
	The features below \SI{100}{\kelvin} in the $M_\text{R}(T)$ curves of the Mn-deficient samples cannot be captured with this two-sublattice model. The atypical drop in the $M_R$ between 5~K and 100~K is likely due to domain relaxation as the anisotropy for these samples is smaller that the $x>1$ samples, as seen in Fig.~\ref{fig:hyst}. For these samples, no phases other than the \ce{Ni_{2}In}-type $B8_2$ were observed in XRD, and therefore it is unlikely that a secondary magnetic phase is contributing to the magnetic signal. We therefore limit the fit for the $x = 0.8$ and $0.9$ samples to temperatures above $T = 100$~K, where the two-sublattice model is able to capture the shape of the $M-T$ data. 
	
	Figure~\ref{fig:Jex}(a) show the fitted moments on the $A$- and $B$-sublattices compared to the same moments estimated from the defect model of Fig.~\ref{fig:disorder}. The mean-field values follow the same trend as the DFT-based model with comparable values. However, the B-sublattice moments from the DFT-based model for the two $x<1$ samples are smaller than would allow for V-type compensated ferrimagnetism. For these compositions, Table \ref{tbl:ex_Mn<0} shows that a non-zero $\delta$ is necessary to create a ferrimagnetic sample. The reason why we have added disorder between the 2a and 2d sites is because DFT shows that Mn$_{2d}$ is substantially larger than Mn$_{2c}$, although $\delta \simeq 0.17$ obtained from a fit to $m_{sat}$ does not create a 2d moment that is large enough.
	
	Figure~\ref{fig:Jex}(b) shows that the exchange constants between the $A$ and $B$ sublattices is small for $x<1$ but is antiferromagnetic, consistent with the presence of \ce{Mn_{2c}} defects dominating the inter-sublattice interaction. With increasing Mn concentration, $\mathcal{J_{AB}}$ increases as expected from the increase in \ce{Mn_{2c}} defects inferred from the DFT-based model.    In contrast, the intra-sublattice interactions are ferromagnetic at low compositions, but reverse sign above $x \simeq 1.4$.
	
	\begin{figure}[!h]
		\centering
		\includegraphics{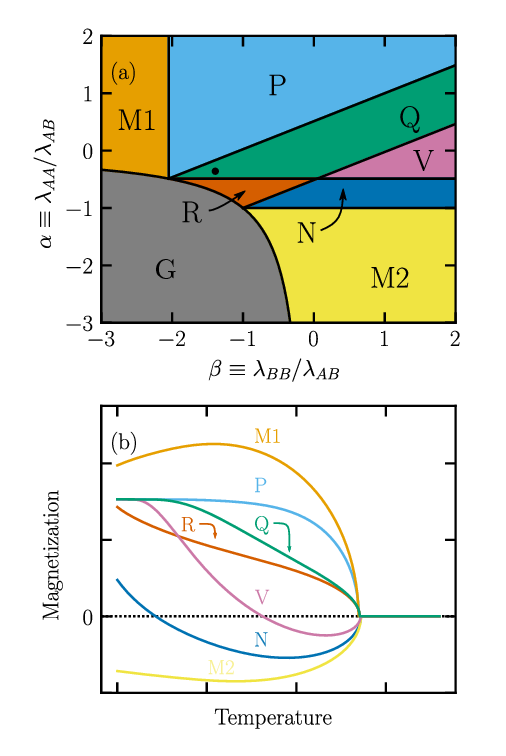}
		\caption{(Color online) The partitioning of molecular field parameter space for a two-sublattice ferrimagnet. Boundaries in (a) were calculated for values of $m_A$ and $m_B$ obtained for \ce{Mn_{2.5}CoGe}. Region G is paramagnetic. The exchange parameters for \ce{Mn_{2.5}CoGe} are shown by the black dot in the Q-type region.  While the precise boundaries in the phase diagrams vary with Mn composition, they remain qualitatively the same for all $x$. Representative $M_\text{R}(T)$ curves for each region are provided in (b).}
		\label{fig:phased}
	\end{figure}
	The evolution in exchange parameters can be mapped onto N\'eel's general ferrimagnetic phase diagram after accounting for the difference between the A- and B-sublattice moments \cite{Neel1948, Smart1966}. The phase diagram is reproduced in Fig.~\ref{fig:phased}~(a) where each of the colored regions in the $\alpha \equiv -\lambda_{AA}/\lambda_{BB}$ and $\beta \equiv -\lambda_{BB}/\lambda_{AB}$ parameter space corresponds to a different shape for $M(T)$, as shown in Fig.~\ref{fig:phased}~(b). The grey region labelled G is paramagnetic for all finite temperatures. 
	The $x = 2.5$ sample resides in the Q-region, near the P-Q  phase boundary, as shown by the black point	. As $x$ decreases, the reduction of Mn on the 2c-sites decreases $\lambda_{AB}$ and increases in the intra-site exchange coupling, which drives the material towards the Q-V boundary and leads to a straightening of the $M-T$ curve at intermediate temperatures in Fig.~\ref{fig:M-T}. This trend continues for $x$ below $x \simeq 1$, and pushes the system into the V-region where compensated ferrimagnetism is observed.
	
	\section{Conclusion}
	Sputtered \ce{Mn_{x}CoGe} compounds formed a metastable \ce{Ni_{2}In}-type structure over the entire compositional range $0.8 \leq x \leq 2.5$ explored in this study. The unexpected ferrimagnetic behavior is explained by the presences of Mn anti-site defects on the 2c/2d-sites. DFT calculations show that these Mn defects are antiferromagnetically coupled to the Mn on the 2a-sites. An atomic model of the distribution of defects in the unit cell using the DFT predicted values explains the general trend in the variations in the saturation magnetization with composition. We provide supporting evidence for the ferrimagnetism with mean-field modeling that both captures variations in the shape of the $M(T)$ curves and the trends in the size of the sublattice moments that follow the DFT-based model. The analysis demonstrates that by increasing the concentration of Mn anti-site defects, the inter-site becomes increasingly antiferromagnetic and the intra-site coupling changes sign, which drives the ferrimagnetism from V-type to Q-type. 
	
	This work suggests the possibility of controlling the ferrimagnetism through defect engineering to generate compensated ferrimagnets in alloys that would otherwise be ferromagnetic. Interest in ferrimagnetism has been revived with the discovery of ultrafast dynamics at the angular momentum compensation point \cite{Wangsness:1953pr, LeCraw:1965jap, Binder:2006prb, Stanciu:2006prb}. Such dynamics could be  valuable in applications for spintronics \cite{Ivanov:2019ltp}, complementary to approaches proposed for devices based on antiferromagnets.
	
	\section{Acknowledgments}
	We would like to thank Jeff Dahn for use of the sputtering machine, as well as Andrew George and Michel Johnson for technical assistance with XRD and PPMS measurements. We also wish to thank James Brenan for the use of the LA-ICP-MS and Erin Keltie for assistance in the collection and analysis of the data. Thank you to Ulrich R\"o\ss ler for helpful discussion about DFT and Cameron Rudderham and Andrey Zelenskiy for insightful conversations.
	\vfill{}
	
	\bibliographystyle{apsrev}

\end{document}